\documentclass[11pt,a4paper]{article}
\usepackage{amsmath}
\usepackage[mathcal]{eucal}
\usepackage{latexsym}
\usepackage{amssymb}
\begin{titlepage}
\title{Instanton representation of Plebanski gravity. Application to Bianchi Type A metrics}
\author{Eyo Eyo Ita III}
\medskip
\input amssym.def
\input amssym.tex

\def \in{\indent}

\begin{document}
\maketitle
\bigskip
\centerline{Department of Applied Mathematics and Theoretical Physics} 
\smallskip
\centerline{Centre for Mathematical Sciences, University of Cambridge, Wilberforce Road}
\smallskip
\centerline{Cambridge CB3 0WA, United Kingdom}
\smallskip
\centerline{eei20@cam.ac.uk} 

\bigskip  
                
\begin{abstract}
Using the instanton representation method we construct a general solution for GR in the spatially homogeneous case restricted to diagonal variables.  This paper provides a testing ground and physical intuition for many of the salient features of general relativity which it is suggested should be preserved in the full theory.
\end{abstract}
\end{titlepage}

\section{Introduction}

In \cite{EYOITA} a new formulation of general relativity has been presented, named the instanton representation of Plebanski gravity.  The basic variables are a $SO(3,C)$ gauge connection $A^a_{\mu}$ and a 3 by 3 matrix $\Psi_{ae}$ which takes its values in two copies of $SO(3,C)$.  In this approach one implements the initial value constraints of general relativity in conjunction with application of a Hodge duality condition to the curvature of $A^a_{\mu}$.  As a consistency condition on this formalism, one requires that the 3-metric $h_{ij}$ determined using the constraint solutions must be equal to the 3-metric defined by the Hodge duality condition.  This amounts to a dynamical condition which evolves the initial data forming the constraint solutions from an initial to a final spatial hypersurface.  The constraint solutions can be classified according to the Petrov type of spacetime, which depends on the multiplicity of eigenvalues of $\Psi_{ae}$ (See e.g. \cite{MACCALLUM} and \cite{PENROSERIND}).\par
\indent
In this paper we apply the instanton representation method to the construction of minisuperspace solutions.  Specifically, we will require all variables to be spatially homogeneous.  This is in line with the standard approaches to GR where one acquires intuition about the full theory from simpler models.  In this paper we highlight some of the salient features from minisuperspace which could be preserved in the full theory.  The organization of this paper is as follows.  In section 2 we provide a self-contained account of some of the relevant results from the instanton representation \cite{EYOITA}.  Section 3 specializes these results to the spatially homogeneous case, and constructs a general solution for the case of diagonal variables.  Sections 4, 5 and 6 construct various minisuperspace solutions for GR, some well-known and others perhaps not as well-known in the literature.  These solutions are labelled essentially by two eigenvalues of $\Psi_{ae}$ and the lapse function $N$, with the connection $A^a_i$ totally eliminated.  In this sense the instanton representation attempts to bring one from the nonmetric back into the metric study of general relativity.

\section{Ingredients for the instanton representation method}

The dynamical variables in the instanton representation of Plebanski gravity are a $SO(3,C)$-valued gauge connection $A^a_{\mu}$ and a 3 by 3 complex matrix $\Psi_{ae}\in{SO}(3,C)\otimes{SO}(3,C)$.\footnote{For index conventions we use lower case symbols from the beginning of the Latin alphabet $a,b,c,\dots$ to denote internal $SO(3,C)$ indices, and from the middle $i,j,k,\dots$ for spatial indices.  Spacetime indices are denoted by $\mu$}  The variables are subject to the following constraints on each 3-dimensional spatial hypersurface $\Sigma$
\begin{eqnarray}
\label{VALUE}
\textbf{w}_e\{\Psi_{ae}\}=0;~~\epsilon_{dae}\Psi_{ae}=0;~~\Lambda+\hbox{tr}\Psi^{-1}=0,
\end{eqnarray}
\noindent
where $\Lambda$ is the cosmological constant.\footnote{The constraints in (\ref{VALUE}) are respectively the Gauss' law, diffeomorphism and Hamiltonian constraints.  These constraints were also written down by Capovilla, Dell and Jacobson in the context of the initial value problem \cite{CAP}.}  We require that $(\hbox{det}\Psi)\neq{0}$, which means that the eigenvalues $\lambda_1$, $\lambda_2$ and $\lambda_3$ of $\Psi_{ae}$ must be nonvanishing.  The first 
equation of (\ref{VALUE}) is defined as
\begin{eqnarray}
\label{GOOSE}
\textbf{w}_e\{\Psi_{ae}\}=\textbf{v}_e\{\Psi_{ae}\}+C_{be}\bigl(f_{abf}\delta_{ge}+f_{ebg}\delta_{af}\bigr)\Psi_{fg}=0,
\end{eqnarray}
\noindent
where $f_{abc}$ are the SO(3) structure constants, and we have defined the vector fields $\textbf{v}_a$ and a magnetic helicity density matrix $C_{ae}$ given by
\begin{eqnarray}
\label{VALUE1}
\textbf{v}_a=B^i_a\partial_i;~~C_{ae}=A^a_iB^i_e.
\end{eqnarray}
\noindent
In (\ref{VALUE1}) we have defined the magnetic field $B^i_a$, which we assume to have nonvanishing determinant, as
\begin{eqnarray}
\label{MAGNET}
B^i_a=\epsilon^{ijk}\partial_jA^a_k+{1 \over 2}\epsilon^{ijk}f_{abc}A^b_jA^c_k.
\end{eqnarray}
\noindent
These variables define a spacetime metric $g_{\mu\nu}$, written in 3+1 form, as
\begin{eqnarray}
\label{METRICAS}
ds^2=-N^2dt^2+h_{ij}{\omega}^i\otimes{\omega}^j,
\end{eqnarray}
\noindent
where $h_{ij}$ is the spatial 3-metric with one forms ${\omega}^i=dx^i+N^idt$, where $N^{\mu}=(N,N^i)$ are the lapse function and shift vector.  The 3-metric $h_{ij}$ is constructed from the constraint solutions, and is given by
\begin{eqnarray}
\label{MEETRIC}
(h_{ij})_{Constraints}=(\hbox{det}\Psi)(\Psi^{-1}\Psi^{-1})^{ae}(B^{-1})^a_i(B^{-1})^e_j(\hbox{det}B),
\end{eqnarray}
\noindent
where $\Psi_{ae}$ and $A^a_i$ are solutions to (\ref{VALUE}).  The constraints (\ref{VALUE}) do not fix $N^{\mu}$, and make use only of the spatial part of the 
connection $A^a_{\mu}$.\par
\indent
From the 4-dimensional curvature $F^a_{\mu\nu}$ one can construct the following object $c_{ij}$, given by
\begin{eqnarray}
\label{CONSTRU}
c_{ij}=F^a_{0i}(B^{-1})^a_j;~~c\equiv\hbox{det}(c_{(ij)}).
\end{eqnarray}
\noindent
The separation of $c_{ij}$ into symmetric and antisymmetric parts defines a 3-metric $(h_{ij})_{Hodge}$ and a shift vector $N^i$, given by
\begin{eqnarray}
\label{CONSTRU1}
(h_{ij})_{Hodge}=-{{N^2} \over c}c_{(ij)};~~N^i=-{1 \over 2}\epsilon^{ijk}c_{jk}.
\end{eqnarray}
\noindent
Equation (\ref{CONSTRU1}) arises from the Hodge duality condition, and is a dynamical statement implied by the instanton representation \cite{EYOITA}.  Equations (\ref{CONSTRU1}) and (\ref{MEETRIC}) are 3-metrics constructed using two separate criteria, and as a consistency condition must be required to be equal to one another.  This is the basic feature of the instanton representation method in constructing GR solutions, which enables us 
to also write (\ref{METRICAS}) as
\begin{eqnarray}
\label{CANBE313}
ds^2=-N^2\bigl(dt^2+{1 \over c}c_{(ij)}\bigl(dx^i-{1 \over 2}\epsilon^{imn}c_{mn}dt)(dx^j-{1 \over 2}\epsilon^{jrs}c_{rs}dt)\bigr).  
\end{eqnarray}
\noindent
Since $\Psi_{ae}$ is a nondegenerate complex matrix by supposition, then it is diagonalizable when there are three linearly independent eigenvectors.  This enables us to classify solutions according to the Petrov type of the self-dual Weyl tensor $\psi_{ae}$.  The matrix $\psi_{ae}$ is symmetric and traceless, and related to $\Psi_{ae}$ in the following way
\begin{eqnarray}
\label{THEWAY}
\Psi^{-1}_{ae}=-{\Lambda \over 3}\delta_{ae}+\psi_{ae}.
\end{eqnarray}
\noindent
So for this paper we assume that $\Psi_{ae}$ is invertible, which requires the existence of three linearly independent eigenvectors.  Hence, the results of this paper are limited to Petrov Types I, D and O.  For each such $\Psi_{ae}$, combined with a connection $A^a_i$ solving the constraints (\ref{VALUE}), the Hodge duality condition (\ref{CONSTRU1}) should yield a metric in (\ref{METRICAS}) solving the vacuum Einstein equations. 

\section{Spatially homogeneous connection}

Let us now consider a spatially homogeneous connection $A^a_{\mu}=A^a_{\mu}(t)$, whose components depend only on time.  The Hamiltonian and diffeomorphism constraints are given by
\begin{eqnarray}
\label{DEEFEEO}
\Lambda+{1 \over {\lambda_1}}+{1 \over {\lambda_2}}+{1 \over {\lambda_3}}=0;~~\epsilon_{dae}\Psi_{ae}=0,
\end{eqnarray}
\noindent
which constrain $\Psi_{ae}$ and not $A^a_i$.  Since all spatial derivatives of $A^a_i$ are zero minisuperspace, then this simplifies the following quantities
\begin{eqnarray}
\label{NEEDEDFOR}
B^i_a=(\hbox{det}A)(A^{-1})^i_a;~~C_{ae}=\delta_{ae}(\hbox{det}A);~~\textbf{v}_a=(\hbox{det}A)(A^{-1})^i_a\partial_i.
\end{eqnarray}
\noindent
On account of the antisymmetry of the structure constants $f_{abc}$, then the Gauss' law constraint simplifies to
\begin{eqnarray}
\label{NEEDEDFOR1}
\textbf{w}_e\{\Psi_{ae}\}=(\hbox{det}A)\bigl((A^{-1})^i_e\partial_i\Psi_{ae}+f_{agf}\Psi_{fg}\bigr)=0.
\end{eqnarray}
\noindent
The second equation of (\ref{DEEFEEO}) causes the second term in brackets in (\ref{GOOSE}) to vanish.  Then defining $k_e\equiv(A^{-1})^i_e\partial_i$, the Gauss' law constraint reduces to $k_e\{\Psi_{ae}\}=0$, which is the statement that $\Psi_{ae}$ must be transverse to the differential operator $k_e$.\footnote{Note for $\Psi_{ae}$ spatially constant, Gauss' law trivializes to $0=0$.  It appears naively that if $\Psi_{ae}$ is allowed to have spatial dependence, then Gauss' law becomes a physical statement concerning the propagation of gravitational waves.  However, the consistency condition (\ref{THEREQUIREMENT}) rules this out as a possibility.}
Define the projection operator $P_{fg}=\delta_{fg}-k^{-2}k_fk_g$.  That this is a projection operator follows from the identities
\begin{eqnarray}
\label{PROJECTOR}
P_{fg}P_{gh}=P_{fh};~~P_{fg}k_g=0.
\end{eqnarray}
\noindent
The general solution is then given by $\Psi_{fg}=P_{fa}P_{ge}\Psi_{ae}$ with $\hbox{tr}\Psi^{-1}=\Lambda$.  This signifies the existence of gravitational waves with respect to the background connection $A^a_i$.  From the 
solution to (\ref{VALUE}), one can construct the spatial 3-metric in accordance with (\ref{MEETRIC}), given by
\begin{eqnarray}
\label{NEEDEDFOR2}
h_{ij}=(\hbox{det}\Psi)(\Psi^{-1}\Psi^{-1})^{ae}A^a_iA^e_j.
\end{eqnarray}
\par
\indent
Having extracted the physical content from the spatial components of the variables, we move on now to the temporal components.  First we construct $c_{ij}$, given by
\begin{eqnarray}
\label{THEFOLLOW}
c_{ij}=(B^{-1})^a_iF^a_{0j}=(\hbox{det}A)^{-1}A^a_i(\dot{A}^a_j-f^{abc}A^b_jA^c_0)\nonumber\\
=(\hbox{det}A)^{-1}\bigl(A^a_{(i}\dot{A}^a_{j)}+A^a_{[i}\dot{A}^a_{j]}\bigr)-\epsilon_{ijk}(A^{-1})^c_kA^c_0,
\end{eqnarray}
\noindent
which separates the symmetric from the antisymmetric parts.  The symmetric part of (\ref{THEFOLLOW}) is a total time derivative given by
\begin{eqnarray}
\label{THEFOLLOW1}
c_{(ij)}=(\hbox{det}A)^{-1}{d \over {dt}}(A^a_{(i}A^a_{j)}),
\end{eqnarray}
\noindent
and the antisymmetric part of $c_{ij}$ defines the shift vector $N^i$, given by
\begin{eqnarray}
\label{THEFOLLOW2}
N^k={1 \over 2}\bigl(\epsilon^{ijk}A^a_j\dot{A}^a_k-(A^{-1})^k_aA^a_0\bigr).
\end{eqnarray}
\noindent
The first term of (\ref{THEFOLLOW2}) has the interpretation of a kind of orbital angular momentum in the space of connections.  For example, in the isotropic case $A^a_i=\delta^a_ia$, where $a=a(t)$ is a function only of time, this term vanishes and the second term reduces to $n^k={1 \over a}A^k_0$.  This provides a convenient physical intrpretation of the shift vector $N^i$ as being directly correlated to the temporal component of the connection $A^a_0$, both being gauge degrees of freedom in their respective theories.\par
\indent
Since the Gauss' law constraint is automatically satisfied in minisuperpsace, then the entire solution reduces to the requirement that (\ref{MEETRIC}), with $\Psi_{ae}$ constrained completely by the Hamiltonian 
constraint, be equal to (\ref{CONSTRU1}).  The latter equation is metric based on the Hodge duality condition from the instanton representation on-shell, which leads to the relation
\begin{eqnarray}
\label{THEREQUIREMENT}
(\hbox{det}\Psi)(\Psi^{-1}\Psi^{-1})^{ae}A^a_iA^e_j=-\Bigl({{N^2} \over {c\hbox{det}A}}\Bigr)\bigl(\dot{A}^e_i-f^{bce}A^b_0A^c_i\bigr)A^e_j.
\end{eqnarray}
\noindent
Equation (\ref{THEREQUIREMENT}) is a first-order time evolution equation for the spatial connection $A^a_i$, which determines the manner of evolution of the physical degrees of freedom in $\Psi_{ae}$.  In this way one may eliminate the connection to construct a metric explicictly in terms of the physical degrees of freedom from (\ref{DEEFEEO}).  In this sense the instanton representation, while starting out as a theory of the connection $A^a_{\mu}$ and the matrix $\Psi_{ae}$, on-shell will produce solutions explicitly in terms of the physical degrees of freedom of the latter.

\subsection{Minisuperspace with diagonal CDJ matrix}
To set the stage for this paper we will establish the necessary ingredients for comparison with the usual Bianchi Type A models, restricted to diagonal variables, for simplicity.  Let us consider a spatially homogeneous 
and diagonal $\Psi_{ae}$ and a diagonal connection, as in
\begin{displaymath}
A^a_i=
\left(\begin{array}{ccc}
a_1 & 0 & 0\\
0 & a_2 & 0\\
0 & 0 & a_3\\
\end{array}\right)
;~~\Psi_{ae}=
\left(\begin{array}{ccc}
\lambda_1 & 0 & 0\\
0 & \lambda_2 & 0\\
0 & 0 & \lambda_3\\
\end{array}\right)
,
\end{displaymath}
\noindent
\noindent
where the eigenvalues satisfy the Hamiltonian constraint relation
\begin{eqnarray}
\label{CONSISTENCY4}
\Lambda+{1 \over {\lambda_1}}+{1 \over {\lambda_2}}+{1 \over {\lambda_3}}=0.
\end{eqnarray}
\noindent
So based entirely on the initial value constraints, in this case equation (\ref{CONSISTENCY4}), one uses (\ref{NEEDEDFOR2}) to directly construct the 3-metric
\begin{displaymath}
(h_{ij})_{Constraints}=
\left(\begin{array}{ccc}
{{\lambda_2\lambda_3(a_1)^2} \over {\lambda_1}} & 0 & 0\\
0 & {{\lambda_3\lambda_1(a_2)^2} \over {\lambda_2}} & 0\\
0 & 0 & {{\lambda_1\lambda_2(a_3)^2} \over {\lambda_3}}\\
\end{array}\right)
.
\end{displaymath}
\noindent
The magnetic field $B^i_a$ and the temporal components of the curvature $F^a_{0i}$ are given in matrix form by
\begin{displaymath}
B^i_a=
\left(\begin{array}{ccc}
a_2a_3 & 0 & 0\\
0 & a_3a_1 & 0\\
0 & 0 & a_1a_2\\
\end{array}\right)
;~~F^a_{0i}=
\left(\begin{array}{ccc}
\dot{a}_1 & N^3 & -N^2\\
-N^3 & \dot{a}_2 & N^1\\
N^2 & -N^1 & \dot{a}_3\\
\end{array}\right)
,
\end{displaymath}
\noindent
which determines the Hodge duality condition.  This fixes the symmetric part of $c_{ij}$,
\begin{displaymath}
c_{(ij)}=
\left(\begin{array}{ccc}
\dot{a}_1(a_2a_3)^{-1} & 0 & 0\\
0 & \dot{a}_2(a_3a_1)^{-1} & 0\\
0 & 0 & \dot{a}_3(a_1a_2)^{-1}\\
\end{array}\right)
;~~c={{\dot{a}_1\dot{a}_2\dot{a}_3} \over {(a_1a_2a_3)^2}}.
\end{displaymath}
\noindent
to the 3-metric $h_{ij}$, and the antisymmetric part of $c_{ij}$ determines the shift vector
\begin{eqnarray}
\label{SHIFTVECTOR}
N^1={{A^1_0} \over {2a_1}};~~N^2={{A^2_0} \over {2a_2}};~~N^3={{A^3_0} \over {2a_3}}.
\end{eqnarray}
\noindent
Note that the orbital angular momentum term, the first term of (\ref{THEFOLLOW2}), is zero for a diagonal connection.  Note also that there is a direct correlation between the gauge degrees of freedom $N^i$ in the metric theory with the temporal curvature components $A^a_0$ in the gauge theory.\par
\indent
We can now compute $\underline{c}_{(ij)}={1 \over c}c_{(ij)}$, which yields
\begin{displaymath}
\underline{c}_{(ij)}=(a_1a_2a_3)^2
\left(\begin{array}{ccc}
(a_2\dot{a}_2)^{-1}(a_3\dot{a}_3)^{-1} & 0 & 0\\
0 & (a_3\dot{a}_3)^{-1}(a_1\dot{a}_1)^{-1} & 0\\
0 & 0 & (a_1\dot{a}_1)^{-1}(a_2\dot{a}_2)^{-1}\\
\end{array}\right)
.
\end{displaymath}
\noindent
Consistency between this and the spatial metric determined by the initial value constraints, namely between (\ref{MEETRIC}) and (\ref{CONSTRU1}) leads to the conditions
\begin{eqnarray}
\label{CONSISTENCY}
-N^2\Bigl({{\lambda_1} \over {\lambda_2\lambda_3}}\Bigr)=\Bigl({{\dot{a}_2} \over {a_2}}\Bigr)\Bigl({{\dot{a}_3} \over {a_3}}\Bigr);~
-N^2\Bigl({{\lambda_2} \over {\lambda_3\lambda_1}}\Bigr)=\Bigl({{\dot{a}_3} \over {a_3}}\Bigr)\Bigl({{\dot{a}_1} \over {a_1}}\Bigr);\nonumber\\
-N^2\Bigl({{\lambda_3} \over {\lambda_1\lambda_2}}\Bigr)=\Bigl({{\dot{a}_1} \over {a_1}}\Bigr)\Bigl({{\dot{a}_2} \over {a_2}}\Bigr).
\end{eqnarray}
\noindent
Multiplication of all three equations together in (\ref{CONSISTENCY}) yields
\begin{eqnarray}
\label{CONSISTENCY1}
\Bigl({{\dot{a}_1\dot{a}_2\dot{a}_3} \over {a_1a_2a_3}}\Bigr)=\pm{{iN^3} \over {\sqrt{\lambda_1\lambda_2\lambda_3}}}.
\end{eqnarray}
\noindent
Then substitution of (\ref{CONSISTENCY1}) into (\ref{CONSISTENCY}) yields
\begin{eqnarray}
\label{CONSISTENCY2}
{{\dot{a}_1} \over {a_1}}={{iN\sqrt{\lambda_1\lambda_1\lambda_3}} \over {(\lambda_1)^2}};~~
{{\dot{a}_2} \over {a_2}}={{iN\sqrt{\lambda_1\lambda_1\lambda_3}} \over {(\lambda_2)^2}};~~
{{\dot{a}_3} \over {a_3}}={{iN\sqrt{\lambda_1\lambda_1\lambda_3}} \over {(\lambda_3)^2}},
\end{eqnarray}
\noindent
We will see that once equations (\ref{CONSISTENCY2}) are fixed by (\ref{CONSISTENCY4}), the entire freedom of the solution becomes reduced to the choice of the lapse function $N$.

\section{Solutions with constant eigenvalues}

Let us construct a solution where the eigenvalues $\lambda_1$, $\lambda_2$ and $\lambda_3$ are independent of time.  The choice of gauge 
\begin{eqnarray}
\label{CONSTANT}
N=-{{in} \over t},
\end{eqnarray}
\noindent
where $n$ is a constant of mass dimension $[n]=1$ which will be chosen suitably.  Then equation (\ref{CONSISTENCY}) becomes
\begin{eqnarray}
\label{CONSTANT1}
{{da_1} \over {a_1}}={{c_1dt} \over t};~~{{da_2} \over {a_2}}={{c_2dt} \over t};~~{{da_3} \over {a_3}}={{c_3dt} \over t},
\end{eqnarray}
\noindent
where we have defined
\begin{eqnarray}
\label{CONSTANT2}
c_1={{n\sqrt{\lambda_1\lambda_2\lambda_3}} \over {(\lambda_1)^2}};~~
c_2={{n\sqrt{\lambda_1\lambda_2\lambda_3}} \over {(\lambda_2)^2}};~~
c_3={{n\sqrt{\lambda_1\lambda_2\lambda_3}} \over {(\lambda_3)^2}}.
\end{eqnarray}
\noindent
Equation (\ref{CONSTANT1}) integrates to
\begin{eqnarray}
\label{CONSTANT3}
a_1(t)=a_1(0)\Bigl({t \over {t_0}}\Bigr)^{c_1};~~a_2(t)=a_2(0)\Bigl({t \over {t_0}}\Bigr)^{c_2};~~a_3(t)=a_3(0)\Bigl({t \over {t_0}}\Bigr)^{c_3},
\end{eqnarray}
\noindent
where the following relation holds on account of the Hamiltonian constraint (\ref{CONSISTENCY4})
\begin{eqnarray}
\label{CONSTANT4}
\sqrt{c_1}+\sqrt{c_2}+\sqrt{c_3}=n^{1/2}(\lambda_1\lambda_2\lambda_3)^{1/4}\Bigl({1 \over {\lambda_1}}+{1 \over {\lambda_2}}+{1 \over {\lambda_3}}\Bigr)=-\Lambda{n}^{1/2}(\lambda_1\lambda_2\lambda_3)^{1/4}.
\end{eqnarray}
\noindent
The choice $c\sqrt{\lambda_1\lambda_2\lambda_3}={1 \over {\Lambda^2}}$ then yields $\sqrt{c_1}+\sqrt{c_2}+\sqrt{c_3}=1$.\par
\indent
We must next construct the 3-metric, which is given by
\begin{displaymath}
h_{ij}(t)=\lambda_1\lambda_2\lambda_3
\left(\begin{array}{ccc}
(a_1/\lambda_1)^2 & 0 & 0\\
0 & (a_2/\lambda_2)^2 & 0\\
0 & 0 & (a_3/\lambda_3)^2\\
\end{array}\right)
\end{displaymath}
\begin{displaymath}
=\lambda_1\lambda_2\lambda_3
\left(\begin{array}{ccc}
\Bigl({{a_1(0)} \over {\lambda_1}}\Bigr)^2\Bigl({t \over {t_0}}\Bigr)^{2c_1} & 0 & 0\\
0 & \Bigl({{a_2(0)} \over {\lambda_2}}\Bigr)^2\Bigl({t \over {t_0}}\Bigr)^{2c_2} & 0\\
0 & 0 & \Bigl({{a_3(0)} \over {\lambda_3}}\Bigr)^2\Bigl({t \over {t_0}}\Bigr)^{2c_3}\\
\end{array}\right)
.
\end{displaymath}
\noindent
With the choice of initial data
\begin{eqnarray}
\label{CONSTANT5}
a_1(0)=\epsilon\sqrt{{{\lambda_1} \over {\lambda_2\lambda_3}}};~~
a_2(0)=\epsilon\sqrt{{{\lambda_2} \over {\lambda_3\lambda_1}}};~~
a_3(0)=\epsilon\sqrt{{{\lambda_3} \over {\lambda_1\lambda_2}}},
\end{eqnarray}
\noindent
then we obtain the following line element
\begin{eqnarray}
\label{CONSTANT6}
ds^2=-\Bigl({1 \over {\lambda_1\lambda_2\lambda_3\Lambda^2}}\Bigr)t^{-2}\Bigl[dt^2+\epsilon^2\Bigl(\Bigl({t \over {t_0}}\Bigr)^{2c_1}dx^2+
(\Bigl({t \over {t_0}}\Bigr)^{2c_2}dy^2+(\Bigl({t \over {t_0}}\Bigr)^{2c_3}dz^2\Bigr)\Bigr],
\end{eqnarray}
\noindent
where 
\begin{eqnarray}
\label{CONSTANT7}
\epsilon=\pm{1}~(Euclidean~signature);~~\epsilon=\pm{i}~(Lorentzian~signature),
\end{eqnarray}
\noindent
and the constants $c_a$ satisfy the relation
\begin{eqnarray}
\label{CONSTANT8}
\sqrt{c_1}+\sqrt{c_2}+\sqrt{c_3}=1.
\end{eqnarray}
\noindent
There are a few things to note regarding the result. (i) The signature of the spacetime is not determined by the lapse function $N$, but rather by the initial data. (ii) The metric (\ref{CONSTANT6}) satisfies the Hodge-duality condition. (iii) This metric is conformally related to a metric of Kasner type.

\section{Nonconstant eigenvalue solution}
We have oconstructed solutions which are conformally related to the Kasner metric, which is a known solution of conventional GR.  But we would also like to construct Kasner solutions.  Let us look for solutions where the eigenvalues have a common scale, by choosing
\begin{eqnarray}
\label{CONSTANT9}
\lambda_1={{p_1} \over {a_1a_2a_3}};~~\lambda_2={{p_2} \over {a_1a_2a_3}};~~\lambda_3={{p_3} \over {a_1a_2a_3}},
\end{eqnarray}
\noindent
where $p_1$, $p_2$ and $p_3$ are numerical constants of mass dimension $1$ satisfying the Hamiltonian constraint
\begin{eqnarray}
\label{CONSTANT10}
{1 \over {p_1}}+{1 \over {p_2}}+{1 \over {p_3}}=0
\end{eqnarray}
\noindent
for $\Lambda=0$.\footnote{Note for $\Lambda=0$ that the Hamiltonian constraint is invariant under rescaling of the eigenvalues.  It is also possible to consider the $\Lambda\neq{0}$ case, but we will not treat it in the present paper for the purposes of brevity.}  Then from (\ref{CONSISTENCY2}) we have
\begin{eqnarray}
\label{CONSTANT11}
{{\dot{a}_1} \over {a_1}}=iN(a_1a_2a_3)^{1/2}{{\sqrt{p_1p_2p_3}} \over {(p_1)^2}};\nonumber\\
{{\dot{a}_2} \over {a_2}}=iN(a_1a_2a_3)^{1/2}{{\sqrt{p_1p_2p_3}} \over {(p_2)^2}};\nonumber\\
{{\dot{a}_3} \over {a_3}}=iN(a_1a_2a_3)^{1/2}{{\sqrt{p_1p_2p_3}} \over {(p_3)^2}}.
\end{eqnarray}
\noindent
Defining $Q\equiv{a}_1a_2a_3$ and choosing the unit lapse $N=1$, then summation of the three equations of (\ref{CONSTANT11}) gives
\begin{eqnarray}
\label{CONSTANT12}
{{dQ^{-1/2}} \over {dt}}=-{i \over 2}\sqrt{p_1p_2p_3}\Bigl[\Bigl({1 \over {p_1}}\Bigr)^2+\Bigl({1 \over {p_2}}\Bigr)^2+\Bigl({1 \over {p_3}}\Bigr)^2\Bigr].
\end{eqnarray}
\noindent
It will be convenient to make the following definitions
\begin{eqnarray}
\label{CONSTANT13}
\eta=\sqrt{p_1p_2p_3}\Bigl[\Bigl({1 \over {p_1}}\Bigr)^2+\Bigl({1 \over {p_2}}\Bigr)^2+\Bigl({1 \over {p_3}}\Bigr)^2\Bigr];~~
c_a={{2\Bigl({1 \over {p_a}}\Bigr)^2} \over {({1 \over {p_1}}\Bigr)^2+\Bigl({1 \over {p_2}}\Bigr)^2+\Bigl({1 \over {p_3}}\Bigr)^2}}
\end{eqnarray}
\noindent
for $a=1,2,3$.  Note that $c_1+c_2+c_3=2$.  Then (\ref{CONSTANT1}) integrates to
\begin{eqnarray}
\label{CONSTANT14}
a_1(t)=a_1(0)\tau^{-c_1};~~a_2(t)=a_2(0)\tau^{-c_2};~~a_3(t)=a_3(0)\tau^{-c_3}
\end{eqnarray}
\noindent
where we have defined
\begin{eqnarray}
\label{CONSTANT15}
\tau\equiv1-{i \over 2}\sqrt{Q_0}\eta{t}
\end{eqnarray}
\noindent
with $Q_0=Q(0)$.  This significance of this will become clear once we choose compute the 3-metric, given by
\begin{displaymath}
h_{ij}(t)=\Bigl({{p_1p_2p_3} \over {Q_0}}\Bigr)\tau^2
\left(\begin{array}{ccc}
\Bigl({{a_1(0)} \over {p_1}}\Bigr)^2\tau^{-2c_1} & 0 & 0\\
0 & \Bigl({{a_2(0)} \over {p_2}}\Bigr)^2\tau^{-2c_2} & 0\\
0 & 0 & \Bigl({{a_3(0)} \over {p_3}}\Bigr)^2\tau^{2c_3}\\
\end{array}\right)
.
\end{displaymath}
\noindent
The choice of initial data
\begin{eqnarray}
\label{CONSTANT16}
a_1(0)=\epsilon\sqrt{{{p_1Q_0} \over {p_2p_3}}};~~
a_2(0)=\epsilon\sqrt{{{p_2Q_0} \over {p_3p_1}}};~~
a_3(0)=\epsilon\sqrt{{{p_3Q_0} \over {p_1p_2}}}
\end{eqnarray}
\noindent
produces the following line element
\begin{eqnarray}
\label{CONSTANT17}
ds^2=d\tau^2+\epsilon^2\Bigl(\tau^{2-2c_1}dx^2+\tau^{2-2c_2}dy^2+\tau^{2-2c_3}dz^2\Bigr).
\end{eqnarray}
\noindent
As a consistency condition on (\ref{CONSTANT16}), we must require that the product of all three initial values of the connection yield $Q_0$, which leads to the condition $Q_0=\epsilon^6p_1p_2p_3$.\footnote{The nondegeneracy condition $(\hbox{det}B)\neq{0}$ is then the same as the nondegeneracy condittion $(\hbox{det}\Psi)\neq{0}$, which is required for the instanton representation to be equivalent to general relativity.}  Then (\ref{CONSTANT15}) is given by
\begin{eqnarray}
\label{CONSTANT18}
\tau=1-{i \over 2}\epsilon^3(p_1p_2p_3)\Bigl[\Bigl({1 \over {p_1}}\Bigr)^2+\Bigl({1 \over {p_2}}\Bigr)^2+\Bigl({1 \over {p_3}}\Bigr)^2\Bigr]t.
\end{eqnarray}
\noindent
The metric (\ref{CONSTANT17}) starts as the Minkoski metric at $t=0$ and then evolves analogously to a Kasner metric.  Note that (\ref{CONSTANT17}) is in general complex since (\ref{CONSTANT18}) can be complex.  But for real general relativity we must require the 3-metric $h_{ij}$ to be real, and for its reality to be preserved under time evolution.  This leads to the following possibilities depending on the signs of the eigenvalues.\par
\indent
\subsection{Reality conditions}
From the Hamiltonian constraint (\ref{CONSTANT10}) there are two possibilities for the three eigenvalues.  We can either have 1 positive and two negative, or one negative and two positive.  We will see that the reality conditions are intimately linked to the signature of the spacetime.\par
\noindent
Case(i): One positive and two negative real eigenvalues.  Then (\ref{CONSTANT18}) is given by
\begin{eqnarray}
\label{CONSTANT19}
\tau=1-{i \over 2}\epsilon\sqrt{\vert{p_1p_2p_3}\vert}\vert\eta\vert,
\end{eqnarray}
\noindent
which requires that $\epsilon$ be pure imaginary as a necessary and sufficient condition for a real metric.  This case corresponds to Lorentzian signature with possibilities
\begin{eqnarray}
\label{CONSTANT20}
\epsilon=i:~Big~crunch~at~t={2 \over {\eta\sqrt{\vert{p_1p_2p_3}\vert}}};\nonumber\\
\epsilon=-i:~Big~crunch~singularity~avoided.
\end{eqnarray}
\noindent
The reality conditions are intimately connected with the signature of spacetime as we can see.\par
\noindent
Case(ii): One negative and two positive pure imaginary eigenvalues.  Then (\ref{CONSTANT18}) is given by
\begin{eqnarray}
\label{CONSTANT21}
\eta=1-{1 \over 2}\epsilon^3\sqrt{\vert{p_1p_2p_3}\vert}\vert\eta\vert,
\end{eqnarray}
\noindent
which requires that $\epsilon$ be real as a necessary and sufficient condition for a real metric.  This case corresponds to Euclidean signature with possibilities
\begin{eqnarray}
\label{CONSTANT22}
\epsilon=1:~Big~crunch~at~t={2 \over {\eta\sqrt{\vert{p_1p_2p_3}\vert}}};\nonumber\\
\epsilon=-1:~Big~crunch~singularity~avoided.
\end{eqnarray}

\section{Exponentially inflating metric solutions}
We will now consider exponentially inflating metrics with constant eigenvalues.  The Hodge-duality condition (\ref{CONSISTENCY2}) yields
\begin{eqnarray}
\label{CONSTANT23}
{{\dot{a}_1} \over {a_1}}=iN\eta_1;~~{{\dot{a}_2} \over {a_2}}=iN\eta_2;~~{{\dot{a}_3} \over {a_3}}=iN\eta_3,
\end{eqnarray}
\noindent
where we have defined for $a=1,2,3$
\begin{eqnarray}
\label{CONSTANT24}
\eta_a={{\sqrt{\lambda_1\lambda_2\lambda_3}} \over {(\lambda_a)^2}};~~\eta_1+\eta_2+\eta_3=\eta
=\sqrt{\lambda_1\lambda_2\lambda_3}\Bigl[\Bigl({1 \over {\lambda_1}}\Bigr)^2+\Bigl({1 \over {\lambda_2}}\Bigr)^2+\Bigl({1 \over {\lambda_3}}\Bigr)^2\Bigr].
\end{eqnarray}
\noindent
Summation of all three equations of (\ref{CONSTANT23}) yields
\begin{eqnarray}
\label{CONSTANT25}
{{\dot{Q}} \over Q}=iN\eta\longrightarrow{Q}(t)=Q_0e^{i\eta\tau},
\end{eqnarray}
\noindent
where $Q=a_1a_2a_3$ and $\tau=\int{N}dt$.  This yields a 3-metric
\begin{displaymath}
h_{ij}(t)=\lambda_1\lambda_2\lambda_3
\left(\begin{array}{ccc}
\Bigl({{a_1(0)} \over {\lambda_1}}\Bigr)^2e^{i\eta_1\tau} & 0 & 0\\
0 & \Bigl({{a_2(0)} \over {\lambda_2}}\Bigr)^2e^{i\eta_2\tau} & 0\\
0 & 0 & \Bigl({{a_3(0)} \over {\lambda_3}}\Bigr)^2e^{i\eta_3\tau}\\
\end{array}\right)
.
\end{displaymath}
\noindent
With a choice of initial data
\begin{eqnarray}
\label{CONSTANT26}
a_1(0)=\epsilon\sqrt{{\lambda_1} \over {\lambda_2\lambda_3}};~~
a_2(0)=\epsilon\sqrt{{\lambda_2} \over {\lambda_3\lambda_1}};~~
a_1(0)=\epsilon\sqrt{{\lambda_3} \over {\lambda_1\lambda_2}},
\end{eqnarray}
\noindent
then we have $Q_0=\epsilon^3(\lambda_1\lambda_2\lambda_3)^{-1/2}$.  The line element is given by
\begin{eqnarray}
\label{CONSTANT27}
ds^2=-d\tau^2+\epsilon^2\Bigl(e^{iN\eta_1\tau}dx^2+e^{iN\eta_2\tau}dy^2+e^{iN\eta_3\tau}dz^2\Bigr).
\end{eqnarray}
\noindent
The condition fixing the signature, $\epsilon=\pm{1}$ for Lorentzian and $\epsilon=\pm{i}$ for Euclidean signatures respectively, seems independent of the reality conditions.  But the reality condition can be phrased as a condition that the spatial volume element $\sqrt{h}$ be real,  
\begin{eqnarray}
\label{CONSTANT28}
\sqrt{h}=Q_0(\lambda_1\lambda_2\lambda_3)^{1/2}e^{i\eta\tau}=\epsilon^3e^{i\eta\tau}.
\end{eqnarray}
\noindent
So the reality condition on (\ref{CONSTANT28}) leads to the conditions: (i) The lapse function $N$ be pure imaginary, and (ii) That $\epsilon$ also be pure imaginary, which limits us to Lorentzian signature spacetimes.\par
\indent
This is the Anisotropic generalization of the De-Sitter type inflating metric in flat coordinates.  Note that such solutions exist both for $\Lambda=0$ as well as for $\Lambda\neq{0}$.

\section{Discussion}

\indent
The results of this paper have demonstrated some of the salient features of the instanton representation method.  In minisuperspace the Gauss' law constraint is automatically satisfied, leaving remaining the Hamiltonian constraint which is a simple algebraic relation between the Weyl tensor eigenvalues.  The entire time evolution of the instanton representation reduces to a first order differential equation for the connection components.  In this paper we have written down a general solution in the case of a diagonal connection, specializing the results to some known minisuperspace solutions.  It is clear that the connection can be totally eliminated, leaving a solution labelled completely in terms of the eigenvalues $\lambda_f$ (which fix the algebraic classification of the spacetime), and the lapse function $N$ which apparently is freely specifiable.  The reality conditions in the instanton representation method appear to be deeply intertwined with the signature of spacetime as determined by the initial data.  This feature is in marked contrast with the situation in the Ashtekar variables.  It is hoped that the results of this paper demonstrate some of the salient features of general relativity which directly extend to the full theory.

\end{document}